\documentclass[runningheads]{llncs}
\usepackage{graphicx}
\usepackage{amsmath} 
\usepackage{amsfonts}
\usepackage{svg}
\usepackage{url}
\usepackage{floatrow}
\usepackage{booktabs}
\usepackage{subcaption}
\newfloatcommand{capbtabbox}{table}[][\FBwidth]

\begin{document}

\title{Equivariant Spherical Deconvolution: Learning  Sparse Orientation Distribution Functions from Spherical Data}

\titlerunning{Equivariant Spherical Deconvolution}

\author{Axel Elaldi\inst{1}\thanks{These authors contributed equally.} \and
Neel Dey\inst{1}\textsuperscript{*} \and
Heejong Kim\inst{1} \and
Guido Gerig\inst{1}}

\authorrunning{A. Elaldi, et al.}

\institute{Department of Computer Science and Engineering, New York University, USA \\
\email{\{axel.elaldi, neel.dey, heejong.kim, gerig\}@nyu.edu}}

\maketitle 

\begin{abstract}
    We present a rotation-equivariant unsupervised learning framework for the sparse deconvolution of non-negative scalar fields defined on the unit sphere. Spherical signals with multiple peaks naturally arise in Diffusion MRI (dMRI), where each voxel consists of one or more signal sources corresponding to anisotropic tissue structure such as white matter. Due to spatial and spectral partial voluming, clinically-feasible dMRI struggles to resolve crossing-fiber white matter configurations, leading to extensive development in spherical deconvolution methodology to recover underlying fiber directions. However, these methods are typically linear and struggle with small crossing-angles and partial volume fraction estimation. In this work, we improve on current methodologies by nonlinearly estimating fiber structures via unsupervised spherical convolutional networks with guaranteed equivariance to spherical rotation. Experimentally, we first validate our proposition via extensive single and multi-shell synthetic benchmarks demonstrating competitive performance against common baselines. We then show improved downstream performance on fiber tractography measures on the Tractometer benchmark dataset. Finally, we show downstream improvements in terms of tractography and partial volume estimation on a multi-shell dataset of human subjects.
\end{abstract}

\section{Introduction}

Diffusion-weighted MRI (dMRI) measures voxel-wise molecular diffusivity and enables the in vivo investigation of tissue microstructure via analysis of white matter fiber configurations and tractography. Localized profiles of water diffusion can be constructed via multiple directional magnetic excitations, with each excitation direction corresponding to an image volume. Due to partial voluming, voxels with two or more crossing fibers require increased directional sampling \cite{wedeen2000mapping} for reliable resolution of multiple fiber directions. However, higher numbers of directions (also referred to as diffusion gradients) lead to clinically infeasible scanning times. Consequently, a series of reconstruction models \cite{daducci2013quantitative} characterizing voxel-specific diffusivity and enabling fewer gradients have been proposed.

In particular, these reconstruction models seek to estimate a \textit{fiber orientation distribution function} (fODF) \cite{tournier2007robust}: a function on the unit sphere $\mathcal{S}^2$ providing fiber orientation/direction and intensity. The fODF can be obtained via spherical deconvolution of the dMRI signal with a tissue \textit{response function} (analogous to a point spread function for planar images). The fODF model represents tissue micro-structure as a sparse non-negative signal, giving higher precision to fiber estimation. 
The constrained spherical deconvolution (CSD) model \cite{tournier2007robust} has been extended to handle multiple tissue types (e.g., white and grey matter, cerebrospinal fluid) with multiple excitation shells (MSMT-CSD) \cite{jeurissen2014multi}. Recent works aim to recover a sparser fODF, either via dictionary-learning \cite{canales2019sparse} or by using specialized basis functions \cite{yan2018estimating}.
However, these methods may fail to recover difficult micro-structures such as crossing-fibers with small crossing angles.

\begin{figure}[t]
\begin{center}
\includegraphics[width=1\linewidth]{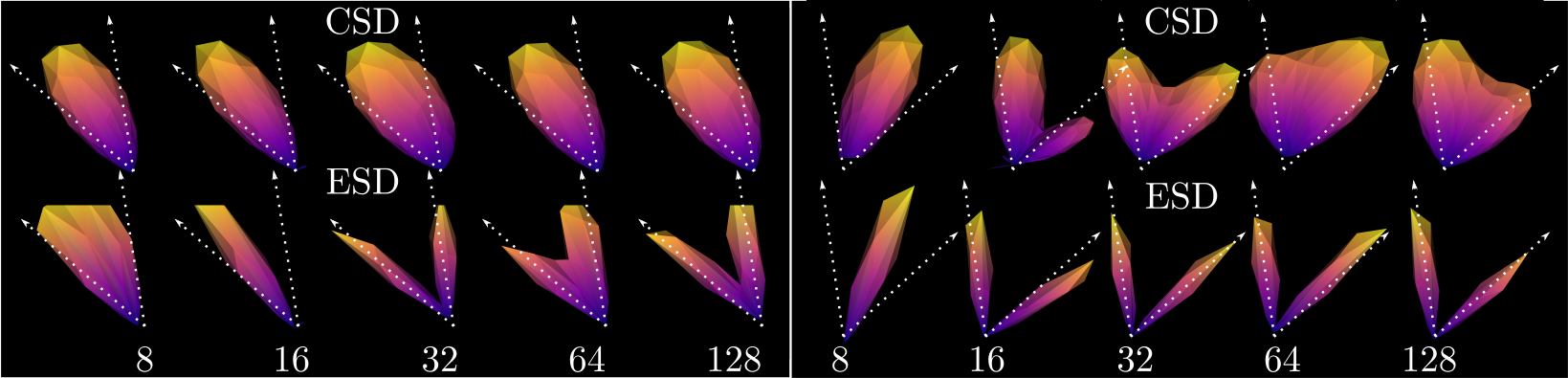}
\end{center}
   \caption{Equivariant Spherical Deconvolution (ESD) can outperform conventional  methods (CSD \cite{jeurissen2014multi}) on the separation of crossing-fibers with small angles (here 33$^{\circ}$ on the left and 45$^{\circ}$ on the right) across a wide range of diffusion gradients.}
\label{fig:csd_esd_sparse_compare}
\end{figure}

Emerging literature demonstrates the utility of deep networks towards learning fODFs. Patel et al. use an autoencoder pretrained for fODF reconstruction as a regularizer for the MSMT-CSD optimization problem \cite{patel2018better}. Nath et al. train a regression network on ground truth fiber orientations acquired via ex-vivo confocal microscopy images of animal histology sections co-registered with dMRI volumes \cite{nath2019deep}. Such an approach is typically impracticable due to the need for ex vivo histological training data. More recently, a series of work \cite{lin2019fast,karimi2020machine,lucena2020using,sedlar2020diffusion} proposes to train supervised deep regression networks directly on pairs of input dMRI signals and their corresponding MSMT-CSD model fits. We argue that such approaches are inherently limited by the quality of the MSMT-CSD solution and show that the underlying deconvolution itself can be improved via an unsupervised deep learning approach. Moreover, none of these learning approaches (with the exception of \cite{sedlar2020diffusion}) are equivariant to spherical rotation. 

In these inverse problems aiming to recover the fODF model, we argue that as the DWI signal lives on the unit sphere, planar convolutional layers may not have the appropriate inductive bias. Standard convolutional layers are constructed for equivariance to planar translation, whereas the analogous operation for spherical signals is rotation. Designing the appropriate form of equivariance for a given network and task is key, as it enables a higher degree of weight sharing, parameter efficiency, and generalization to poses not seen in training data \cite{cohen2016group}. Fortunately, rotation-equivariant spherical convolutional layers for data on $\mathcal{S}^2$ have been proposed \cite{s.2018spherical,perraudin2019deepsphere} with natural applicability to dMRI data.

In this work, we tackle sparse unsupervised fODF estimation via rotation-equivariant spherical convolutional networks. This reformulation is trained under a regularized reconstruction objective and allows for the nonlinear estimation of sparse non-negative fiber structures with incorporation of relevant symmetries and leads to improved performance on a variety of fODF estimation settings for both single-shell and multi-shell data. When ground truth is available via benchmark datasets, we obtain more accurate fiber detection and downstream fiber tractography. For real-world application to humans without ground truth, we show downstream improvements in terms of tractography and partial volume estimation on a real-world multi-shell dataset. Our learning framework is flexible and amenable to various regularizers and inductive priors and is applicable to generic spherical deconvolution tasks. Our code is publicly available at \url{https://github.com/AxelElaldi/equivariant-spherical-deconvolution}.


\begin{figure}[!t]
\centering
\includegraphics[width=\linewidth]{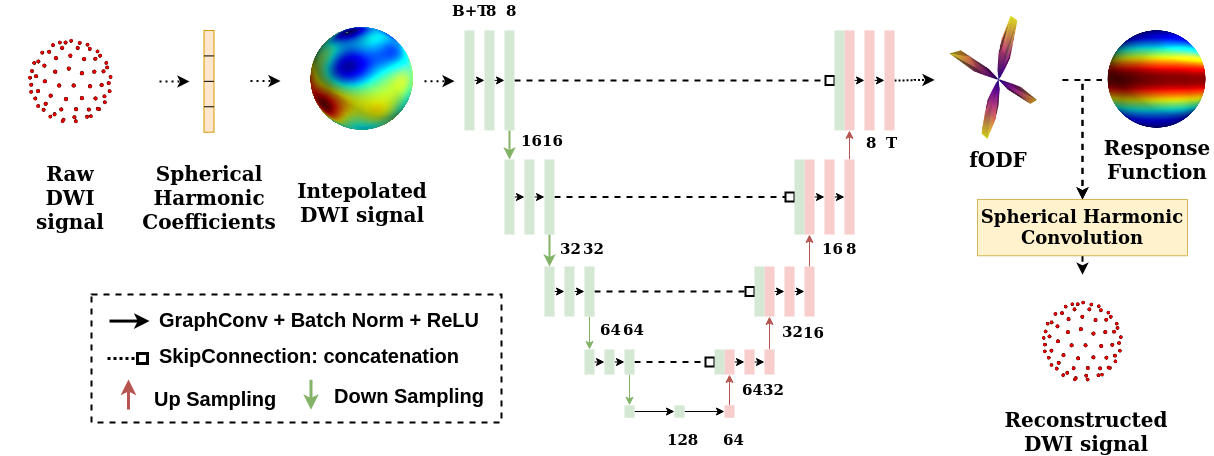}
\caption{\textbf{Framework overview}. The raw DWI signal is interpolated onto a spherical Healpix grid \cite{gorski1999healpix} and fed into a rotation-equivariant spherical U-Net which predicts sparse fiber orientation distribution functions. The architecture is trained under a regularized reconstruction objective.}
\label{fig:overview}
\end{figure}

\section{Methods}

\subsection{Background and Preliminaries}
\noindent\textbf{dMRI Deconvolution.}
The dMRI signal is a function $S:\mathcal{S}^2\rightarrow \mathbb{R}^B$, where $B$ is the number of gradients strengths/sampling shells, with the fODF given as $F:\mathcal{S}^2\rightarrow \mathbb{R}$ such that $S = \mathcal{A}(F)$, where $\mathcal{A}$ is the spherical convolution of the fODF with a response function (RF) $R:\mathcal{S}^2\rightarrow \mathbb{R}^B$. The RF can be seen as the dMRI signal containing only one fiber in the $\mathbf{y}$-axis direction. The spherical convolution is defined as $S(p) = (R * F)(p) = \int_{\mathcal{S}^2} R(P_p^{-1} q) F(q) dq$, where $p,q\in \mathcal{S}^2$ are spherical coordinates and $P_p$ is the rotation associated to the spherical angles of coordinate $p$. We note that the fODF is voxel-dependent and shell-independent with antipodal symmetry and that the RF is voxel-independent, shell-dependent, and rotationally symmetric about the $\mathbf{y}$-axis. Typically, convolution between two $\mathcal{S}^2$ signals yields a signal on $\mathbf{SO}(3)$ (the group of 3D rotations) as the rotation matrix $P_p\in \mathbf{SO}(3)$ is a 3D rotation \cite{s.2018spherical}. Specific to spherical deconvolution, as the RF is symmetric about the $y$-axis, all rotations which differ only by rotation around the $\mathbf{y}$-axis give the same convolution result. Therefore, $P_p$ can be expressed with two angles and the output of the convolution lives on $\mathcal{S}^2$. \\

\begin{figure}[!t]
\begin{center}
\includegraphics[width=0.8\linewidth]{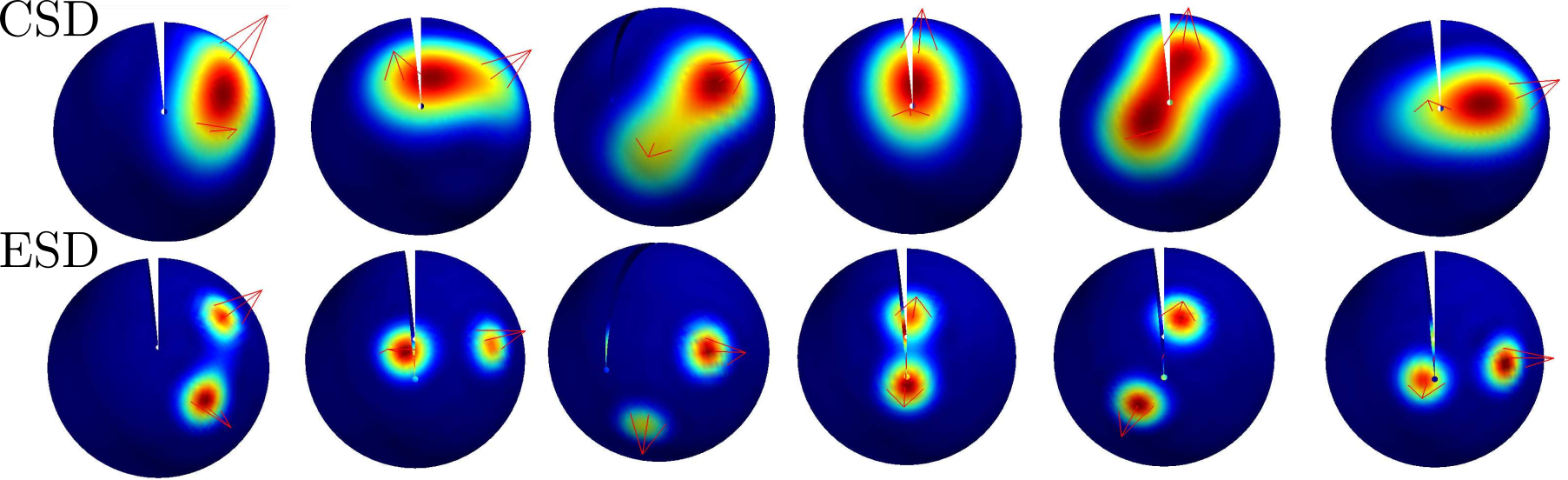}
\end{center}
   \caption{Qualitative synthetic (Sec. \ref{subsec:synth_dataset}) results showing fODF estimation on $128$-gradient 2-fiber samples with red arrows representing ground truth fibers and the heatmap showing model prediction. Row 1: CSD \cite{jeurissen2014multi}, Row 2: ESD (ours).}
\label{fig:csd_esd_fodf_synth_compare}
\end{figure}

\noindent\textbf{Spherical Harmonics.}
We utilize the orthonormal Spherical Harmonics (SH) basis $\{Y_l^m\}_{l\in\mathbb{N}, m\in\{-l,...,l\}}$ $:\mathcal{S}^2\rightarrow \mathbb{R}$ to express square-integrable $f:\mathcal{S}^2\rightarrow \mathbb{R}$ as $f(p)=\sum_{l=0}^{\infty} \sum_{m=-l}^l f_{l,m}Y_l^m(p),$ where $p \in \mathcal{S}^2$, $\{f_{l,m}\}_{l\in\mathbb{N}, m\in\{-l,...,l\}}\in \mathbb{R}$ are the spherical harmonic coefficients (SHC) of $f$. We assume $f$ to be bandwidth limited such that $0\leq l\leq 2l_{max}$. As even degree SH functions are antipodally symmetric and odd degree SH functions are antipodally anti-symmetric, the odd degree SHC of both RF and fODF are null. Moreover, the $m$ order SH functions are azimuthal symmetric only for $m=0$. Thus, RF has only $0$-order SHC. 
Therefore, the dMRI signal $S^b$ for the $b$-th shell is $S^b(p) = \sum_{l=0}^{l_{max}} \sum_{m=-2l}^{2l} \sqrt{\frac{4\pi}{4l+1}}r^b_{2l}f_{2l,m} Y_{2l}^m(p)$ for $l\in\{0,...,l_{max}\}, m\in\{-2l,...,2l\}$ where $p \in \mathcal{S}^2$,  $L=(2l_{max}+1)(l_{max}+1)$ is the number of coefficients, and $\{r^b_{2l,0}\}$ and $\{f_{2l,m}\}$ are the SHC of the RF and fODF, respectively. \\

\noindent\textbf{Matrix Formulation.}
Let the dMRI signal be sampled over $B$ shells for $V$ voxels. For a specific $b$-shell, a set of $n^b$ gradient directions $\{(\theta^b_i, \phi^b_i)\}_{1\leq i \leq n^b}$ is chosen, where $\theta^b_i$ and $\phi^b_i$ are angular coordinates of the $i$th gradient direction. This set gives $n^b$ values $\{S^{b,v}_i\}_{1\leq i \leq n^b}$ for each voxel $v$, where $S^{b,v}_i=S^{b,v}(\theta^b_i, \phi^b_i)$. Let  $\mathbf{S}^b\in\mathcal{M}_{V, n^b}(\mathbb{R})$ be the sampling of the $v$th voxel of the $b$-shell, with the $v$\textsuperscript{th} row being $\{S^{b,v}_i\}_{1\leq i \leq n}$ . Let $\mathbf{Y}^b\in\mathcal{M}_{L,n}(\mathbb{R})$ be the SH sampled on the $i$\textsuperscript{th} gradient of the $b$-shell $\{(\theta^b_i, \phi^b_i)\}$ with its $i$\textsuperscript{th} column being $\{Y_{2l}^m(\theta^b_i, \phi^b_i)\}_{l, m}$. Let $\mathbf{F}\in\mathcal{M}_{V, L}(\mathbb{R})$ be the matrix of the fODF SH coefficients with the $v$\textsuperscript{th} row being the coefficients $\{f_{2l,m}^v\}_{l, m}$ of the $v$\textsuperscript{th} voxel. Finally, let $\mathbf{R}^b\in\mathcal{M}_{L, L}(\mathbb{R})$ be a diagonal matrix, with diagonal elements $\sqrt{\frac{4\pi}{4l+1}}r^b_{2l}$ in blocks of length $4l+1$ for $l\in\{0,...,l_{max}\}$. The diffusion signal can now be written as $\mathbf{S}^b = \mathbf{F}\mathbf{R}^b\mathbf{Y}^b$, with $\mathbf{F}\mathbf{R}^b$ giving the SHC of $S$ and $\mathbf{Y}^b$ transforming SHC into spatial data. \\

\begin{figure*}[!t]
\begin{center}
\includegraphics[width=\linewidth]{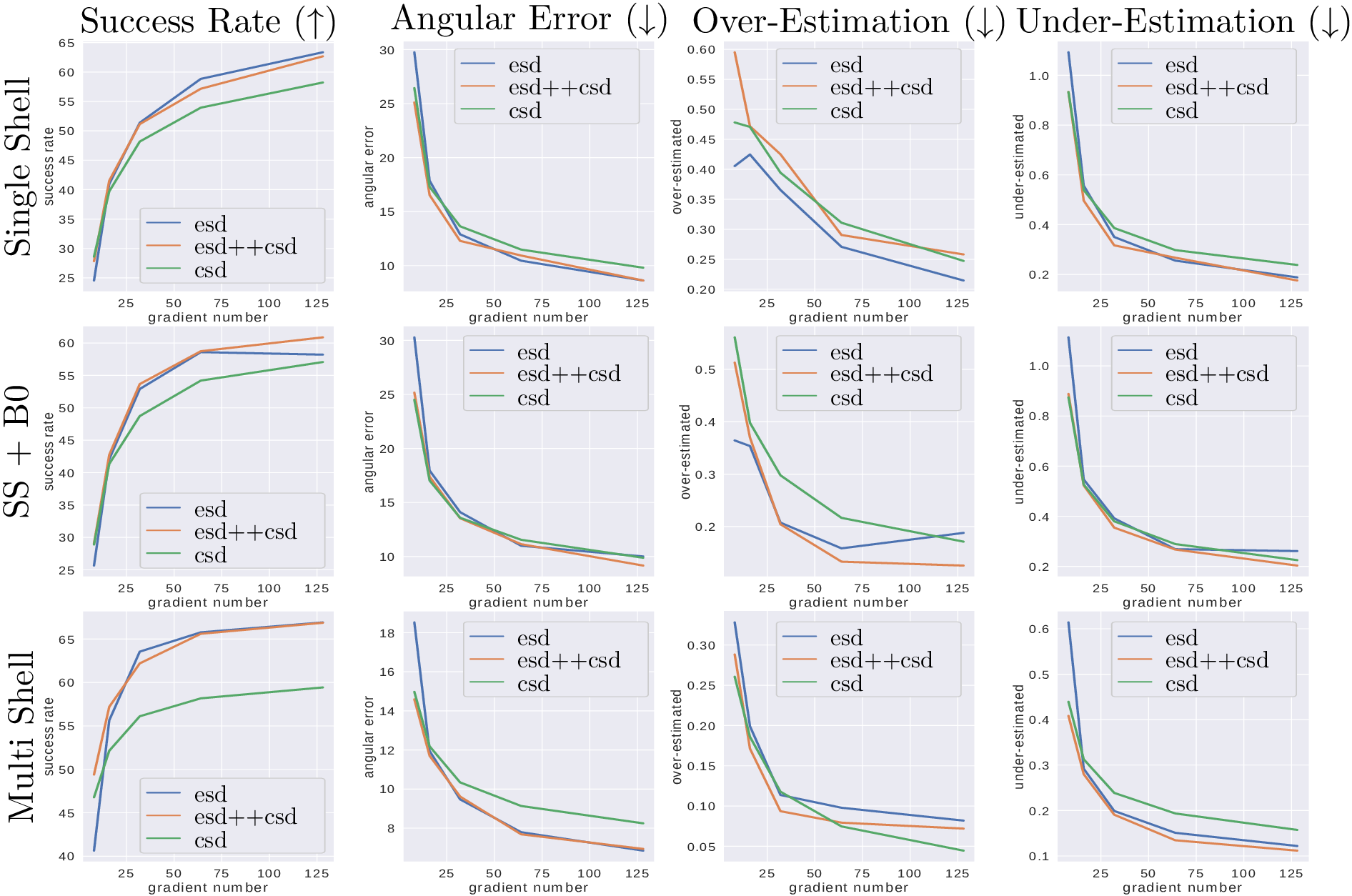}
\end{center}
   \caption{Synthetic results relative to the number of gradients on 3 dMRI settings in terms of Success Rate, Angular Error, Over-estimation, and Under-estimation. Arrows indicate whether lower or higher is better for a given score.}
\label{fig:synth_plot}
\end{figure*}

\noindent\textbf{Multi-Tissue decomposition.}
So far, formalism has been presented for voxels with a single tissue type. In reality, brain tissue comprises multiple components, e.g., white matter (WM), grey matter (GM), and cerebrospinal fluid (CSF). To this end, \cite{jeurissen2014multi} presents a diffusion signal decomposition between WM/GM/CSF such that $S(p) = S_{wm}(p) + S_{gm} + S_{csf}$. GM and CSF are assumed to have isotropic diffusion limiting their spherical harmonic bandwidth to $l_{max}=0$. Thus, $\mathbf{S}^b = \mathbf{F}_{wm}\mathbf{R}_{wm}^b\mathbf{Y}_{wm}^b + (\mathbf{F}_{gm}\mathbf{R}_{gm}^b + \mathbf{F}_{csf}\mathbf{R}_{csf}^b)\mathbf{Y}_{iso}^b$,
where $\mathbf{F}_{gm},\mathbf{F}_{csd}\in\mathcal{M}_{V, 1}(\mathbb{R})$, $\mathbf{R}_{gm},\mathbf{R}_{csd}\in\mathcal{M}_{1, 1}(\mathbb{R})$, and $\mathbf{Y}_{iso}\in\mathcal{M}_{1, n}(\mathbb{R})$.
Simplifying notation, we finally aim to solve the spherical deconvolution optimization problem to find SH coefficients of the $V$ fODFs. Deconvolution methods typically assume the SH coefficients of the RF $\mathbf{R}$ to be known. Thus, the estimated fODF is:
\begin{align}
\label{eq:optim}
    \mathbf{\hat{F}} = \text{argmin}_{\mathbf{F}, \mathbf{FY}\geq 0} ||\mathbf{S}-\mathbf{FRY}||^2 + \lambda Reg(\mathbf{F})
\end{align}
where $Reg$ is a sparsity regularizer on the fODF and $FY\geq 0$ implies a non negativity constraint on the fODF intensities used in \cite{tournier2007robust} with a threshold of $0$.

\subsection{Equivariant Spherical Deconvolution}
Here, we outline how to take an unsupervised deep spherical network towards sparse non-negative fODF estimation, with an overview shown in Figure \ref{fig:overview} and applicability to single-shell single-tissue (SSST), single-shell multi-tissue (SSMT), and multi-shell multi-tissue (MSMT) deconvolution. \\

\noindent\textbf{Graph convolution}
We utilize the rotation-equivariant graph convolution developed in  \cite{perraudin2019deepsphere} due to its improved time complexity over harmonic methods such as \cite{s.2018spherical}. The sphere is discretized into a graph $\mathcal{G}$, such that $f:\mathcal{S}^2\rightarrow\mathbb{R}^B$ is sampled on the $N$ vertices of $\mathcal{G}$ such that the signal becomes a matrix $\mathbf{f}\in \mathcal{M}_{N, B}(\mathbb{R})$. A graph convolution can now be written as $h(\mathbf{L)f} = \sum_{i=0}^P w_i \mathbf{L}^i \mathbf{f}$, where the convolutional filter is fully described by weights $\{w_i\}$, and $\mathbf{L}=\mathbf{D}-\mathbf{A}$ is the graph Laplacian with degree matrix $\textbf{D}$ and adjacency matrix $\textbf{A}$. This graph convolution can be made approximately rotation equivariant following \cite{perraudin2019deepsphere} by fixing the edge weights of the graph and the discretization of the sphere. We use exponential weighting $d_{i,j} = e^{-\frac{||x_i - x_j||_2^2}{\rho^2}}$ if $i$ and $j$ are neighbors and $d_{i,j} = 0$ otherwise, where $i$ and $j$ are vertex indices of the graph, $x_i$ is the coordinate of the $i$-th vertex, and $\rho$ is the average distance between two neighbors. We use hierarchical Healpix sampling \cite{gorski1999healpix} of the sphere to construct our graph. \\

\begin{table*}[!t]
\begin{center}
\begin{tabular}{l|c|c|c|c|c|c}
\toprule
\textbf{Method} & \textbf{\# Tissues} & \textbf{VB ($\uparrow$)} & \textbf{IB ($\downarrow$)} & \textbf{VC ($\uparrow$)} & \textbf{IC ($\downarrow$)} & \textbf{NC ($\downarrow$)} \\
\midrule
CSD & 1 & $\mathbf{22}$  & $117$ & $34.79$ & $65.07$ & $0.15$ \\
ESD & 1 & $\mathbf{22}$  & $123$ & $48.06$ & $51.82$ & $0.12$ \\
ESD+\!\!\!+CSD & 1  & $\mathbf{22}$ & $111$ & $48.80$ & $51.06$ & $0.14$ \\
\hline
CSD & 2 & $21$  & $112$ & $46.81$ & $53.06$ & $0.13$ \\
ESD & 2 & $21$  & $\mathbf{65}$ & $65.12$ & $34.86$ & $0.02$ \\
ESD+\!\!\!+CSD & 2 & $21$ & $72$ & $\mathbf{65.22}$ & $\mathbf{34.77}$ & $\mathbf{0.01}$ \\
\bottomrule
\end{tabular}
\end{center}
\caption{Downstream post-deconvolution tractography results on Tractometer in terms of Valid/Invalid Bundles, Valid/Invalid Connections, and No Connections. Arrows indicate whether lower or higher is better for a given score.}
\label{tab:ismrm_metric}
\end{table*}

\noindent\textbf{Spherical harmonics resampling.}
Real-world DWI acquisition protocols sample diffusion signals over a few dozen to at most a few hundred points with these points not corresponding to Healpix sampling. Therefore, to construct the deconvolution network input, we resample the diffusion signal onto the Healpix grid using spherical harmonics interpolation as illustrated in Figure \ref{fig:overview}.
We thus obtain $\mathbf{S}_{input}\in \mathcal{M}_{V, B, N}(\mathbb{R})$, where $V$ is the number of voxels in a batch, $B$ is the number of shells and $N$ is the number of vertices. In the case the sampling and the frequency bandwidth of the DWI signal are not consistent, this step introduces a non-rotation equivariant operation. However, the model is still equivariant to the set of rotations that permute the sampling gradients. \\

\noindent\textbf{Sparsity.}
Typically CSD methods represent the fODF with spherical harmonics up to degree $8$, implying that the fODF representation cannot approximate a Dirac-like function and thus two close fibers cannot be distinguished. We maintain the spherical harmonic basis, but increase their degree to $20$ for better representation with the ability to separate small crossing-fibers. To ensure the sparsity of the predicted fODF, we forgo $L_1$ regularization as in other sparse reconstruction work \cite{liu2017multichannel} and assume that the fODF follows a heavy-tailed Cauchy distribution and regularize towards it as $Reg(\mathbf{F}) = \sum_{i=1}^N log(1 + \frac{f_i^2}{2\sigma_c^2})$,
where $N$ is the number of points we estimate the fODF on, $f_i$ is the fODF value on the $i$th sphere pixel and $\sigma_c$ controls the sparsity level of the fODF. \\

\begin{figure*}[t]
\begin{center}
\includegraphics[width=\linewidth]{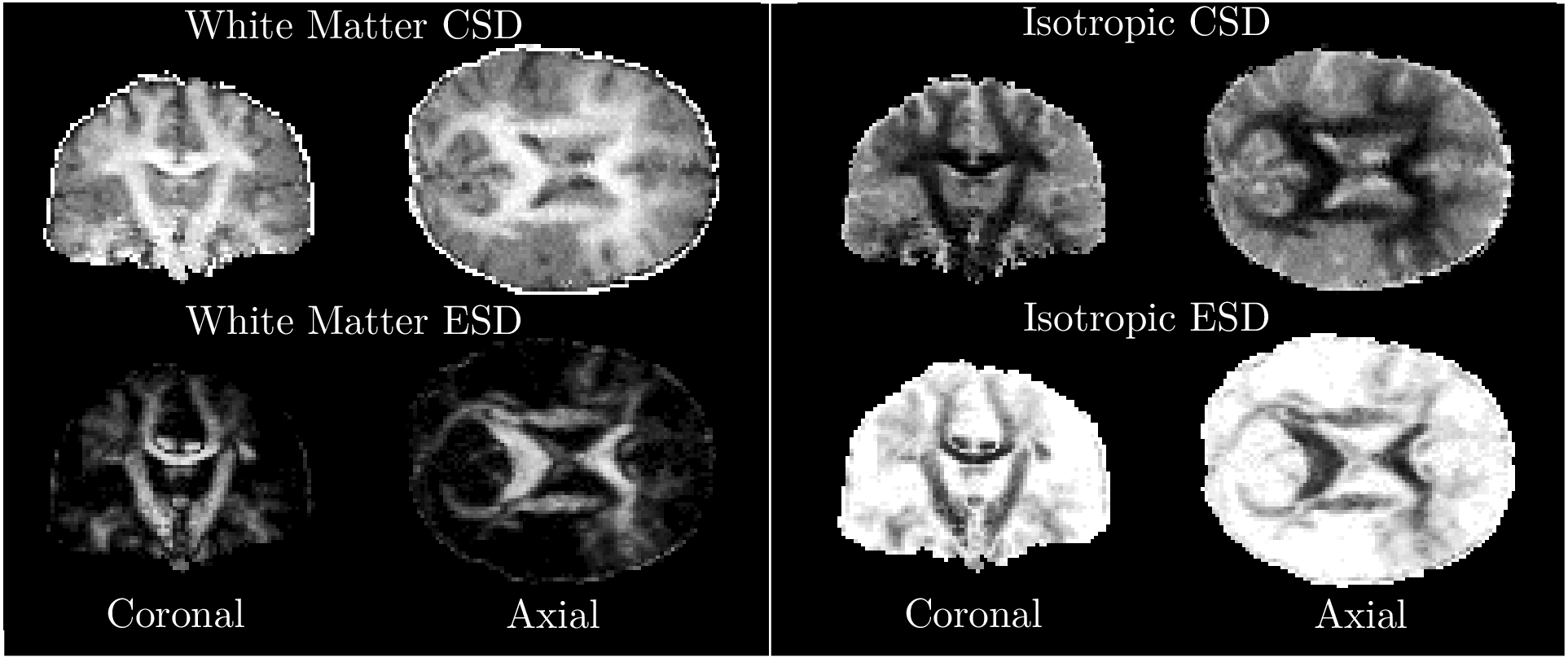}
\end{center}
   \caption{Tractometer partial volume fractions estimated from CSD (row 1) and ESD (row 2), for the WM compartment (col. 1-2) and the isotropic GM and CSF compartments (col. 3-4). ESD returns more accurate localized tissue maps.}
\label{fig:pv_ismrm}
\end{figure*}

\noindent\textbf{Learning Framework.}
An overview of the overall network architecture is described in Fig. \ref{fig:overview}. Response functions for the tissue compartments are pre-calculated with \verb|MRTrix3| \cite{tournier2019mrtrix3}. We input the Healpix-resampled signal into a rotation-equivariant graph convolutional network, following a U-net style architecture. We use max pooling and unpooling over the Healpix grid for down and upsampling, with batch normalization and ReLU nonlinearities following every convolution except for the last layer. For the final pre-fODF layer, we use the Softplus activation for MSMT and ReLU for SSST for increased stability.

The fODF outputs are $T$ signals $\mathbf{O}\in\mathcal{M}_{V,T,N}(\mathbb{R})$ each corresponding to a tissue compartment. The SHCs of the WM fODF are the even SHC degrees of the first output signal. For the isotropic GM and CSF, we take the maximum values of the second and third output signal and use them as the GM and CSF fODF SHC. Finally, the predicted fODF is convolved with the RF to reconstruct the signal $S$, training the network under a regularized reconstruction objective as below. For simplicity, we omit the indices for the multiple tissue compartments.
\begin{align*}
    \mathcal{L}(fODF)=&||S-fODF*RF||_2^2 + \lambda\sum_{i=1}^N log(1 + \frac{fODF_i^2}{2\sigma_c^2}) +||fODF_{fODF<0}||_2^2 
\end{align*}
where the first term represents signal reconstruction, the second corresponds to sparsity as described before, and the third term encouraging non-negativity. As we use a ReLU or Softplus nonlinearity on the network output, the initially estimated fODF is entirely non-negative. However, as we use only the even-order SHC of the fODF for convolution with the RF for reconstruction (the fODF is symmetric), eliminating the odd-order SHC may introduce negative values to the fODF which we suppress using an $L_2$ regularizer on only its negative elements $fODF_{fODF<0}$. We use a batch size of 32 and Adam for minimization with a $10^{-2}$ step size, step-decayed on loss plateau. Depending on the dataset, all networks are found to rapidly converge within 20-30 epochs.

\begin{figure*}[!t]
\begin{center}
\includegraphics[width=\linewidth]{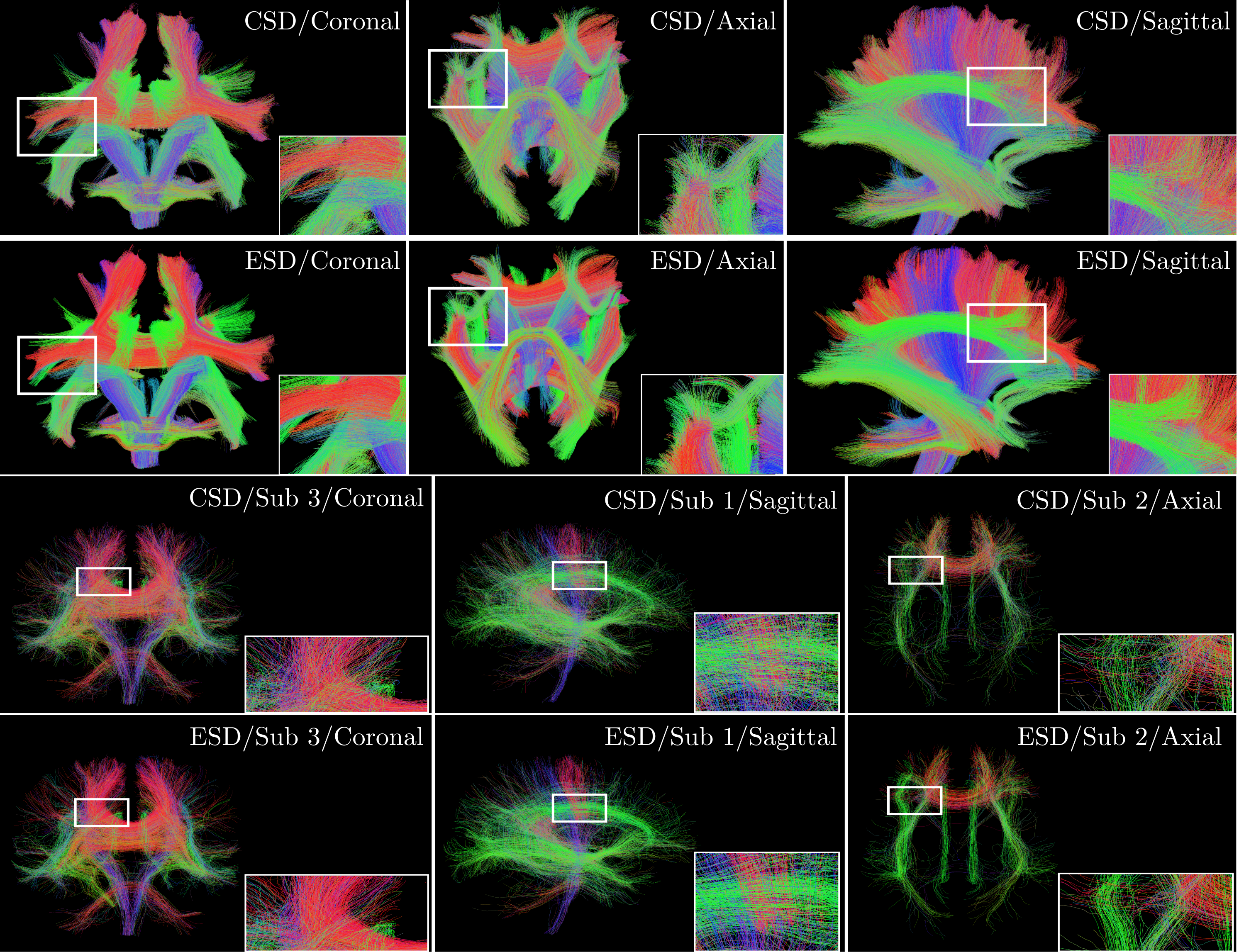}
\end{center}
   \caption{Post-deconvolution tractography for Tractometer (rows 1-2; single-shell) and a human subject (rows 3-4; multi-shell). For both datasets, ESD (rows 2 and 4) demonstrates clearer streamlines with lower noise as opposed to CSD (rows 1 and 3). Readers are encouraged to zoom-in for visual inspection.}
\label{fig:tract_ismrm_human}
\end{figure*}
\section{Experiments}

We benchmark our methodologies across diverse datasets and deconvolution settings including: (1) synthetic data generated from a noisy multi-tensor model where we evaluate SSST, SSMT, and MSMT; (2) the Tractometer dMRI benchmark for SSST and SSMT; (3) an in-vivo human multi-shell dataset (MSMT). We compare our methods against the state-of-the-art CSD implementations available in \verb|MRTrix3| \cite{tournier2019mrtrix3}. We note that ground truth fODFs are not available for Tractometer and the human dataset, motivating our use of surrogate evaluations in terms of downstream utility via tractography and partial volume estimation. Finally, for completeness, we test whether feature engineering by way of concatenating CSD deconvolutions to the network input (denoted as ESD+\!\!\!+CSD) would provide performance improvements.

\subsection{Noisy Synthetic Benchmark} \label{subsec:synth_dataset}
\noindent\textbf{Dataset.} We use the multi-tissue diffusion model from \cite{jeurissen2014multi} to generate a multi-shell multi-tissue dataset. Representative response functions are estimated from a human subject (subject 1, center 1 from \cite{tong2020multicenter}) using \verb|MRTrix|.
We add Rician noise to the simulated signal corresponding to real MR noise. To assess the robustness of the method against the number of gradients and shells, we generate five single-shell datasets with $b$-value $3000 s/mm^2$ and $\{8, 16, 32, 64, 128\}$ diffusion gradients per method. We also generate five multi-shell datasets ($b = \{1000, 2000, 3000\} s/mm^2$) corresponding to $\{8, 16, 32, 64, 128\}$ diffusion gradients per shell. Finally, we generate a single $b=0$ signal. We simulate $1e5$ samples, split into $\{7e4, 1e4, 2e4\}$ for training, validation and testing. \\

\noindent\textbf{Evaluation Scores.} As ground truth microstructure is known, performance evaluation is performed via five scores: (1) The \textit{success rate} measures the ability to correctly estimate the number and location of white matter fibers via the number of voxels corrected processed. A voxel is successfully processed by the model if each ground truth fiber can be match to a predicted fiber and the number of predicted fibers is the same as ground truth fibers. Following \cite{canales2019sparse}, a ground truth fiber is matched to a predicted fiber if it is no further than $25$ degrees away; (2) The \textit{angular error} measures angular distance between a ground truth fiber and the closest predicted fiber; (3) The \textit{overestimatation error} estimates the number of predicted fibers outside of the $25$-degree cone of the ground truth fibers; (4) The \textit{underestimation error} measures the number of ground-truth fibers without predicted fibers in their $25$-degree cone; (5) Finally, we measure the estimated fractional volume of each tissue component (a probability mass function) per voxel in terms of the KL-divergence to the ground truth. For scores 1-4, we use the peak detection algorithm from \cite{canales2019sparse} to predict fiber directions, where the amplitude threshold selection is done on each validation set. \\

\noindent\textbf{Results.} Evaluation scores are presented in Fig. \ref{fig:synth_plot}. Our models consistently improve success rate, angular error and under-estimation over all gradients, shells, and number of tissue decompositions, save for $8$ gradients. These results suggest that sharp fODFs returned by our methods allow for better localization and detection of fibers, decreased undetected fibers, while reducing the number of spurious fibers detected. Fig. \ref{fig:csd_esd_fodf_synth_compare} shows the better capacity of ESD to detect small-angle crossing fibers over CSD. Table \ref{tab:kl} demonstrates better partial volume fraction estimation with our models, leading to better tissue component estimation and better localization of white matter fibers.

\begin{table*}[!t]
\begin{minipage}[b]{0.6\linewidth}
\centering
\begin{tabular}{lcccccc}
\toprule
\multicolumn{2}{c}{\textbf{ }} & \multicolumn{5}{c}{\textbf{Number of Gradients}} \\
\textbf{Method} & \textbf{\#Tissues} & \textbf{8} & \textbf{16} & \textbf{32} & \textbf{64} & \textbf{128} \\
\midrule
CSD & 2 & $0.46$  & $\mathbf{0.40}$ & $0.39$ & $0.39$ & $0.38$ \\
ESD & 2 & $0.39$  & $0.61$ & $0.14$ & $\mathbf{0.1}3$ & $0.15$ \\
ESD+\!\!\!+CSD & 2  & $\mathbf{0.31}$ & $0.66$ & $\mathbf{0.12}$ & $0.15$ & $\mathbf{0.13}$ \\
\hline
CSD & 3 & $0.54$  & $0.44$ & $0.38$ & $0.36$ & $0.35$ \\
ESD & 3 & $\mathbf{0.09}$  & $\mathbf{0.07}$ & $\mathbf{0.06}$ & $\mathbf{0.06}$ & $\mathbf{0.06}$ \\
ESD+\!\!\!+CSD & 3 & $0.10$ & $\mathbf{0.07}$ & $\mathbf{0.06}$ & $\mathbf{0.06}$ & $0.07$ \\
\bottomrule
\end{tabular}
\label{fig:figure1}
\end{minipage}
\hspace{0.02\linewidth}
\begin{minipage}[b]{0.35\linewidth}
\centering
\begin{tabular}{lcccc}
\toprule
\textbf{Method} & \textbf{S. 1} & \textbf{S. 2} & \textbf{S. 3} \\
\midrule
CSD & $3.46$ & $3.10$ & $3.42$ \\
ESD & $1.23$ & $0.98$ & $\mathbf{0.93}$  \\
ESD+\!\!\!+CSD & $\mathbf{0.83}$ & $\mathbf{0.82}$ & $0.99$ \\
\bottomrule
\end{tabular}
\label{fig:figure2}
\end{minipage}
\caption{Partial Volume Fraction estimation in terms of KL divergence to ground truth. \textbf{Left:} synthetic performance relative to the number of gradients. \textbf{Right:} Performance on a multi-shell Human dataset for individual subjects (S1-3).}
\label{tab:kl}
\end{table*}

\subsection{The Tractometer Benchmark} \label{subsec:tractometer}
\noindent\textbf{Dataset.} To assess downstream deconvolution utility, we utilize the ISMRM 2015 Tractometer challenge \cite{maier2017challenge} which provides a realistic single-shell multi-tissue brain phantom with $25$ ground truth fiber-bundles, $32$ diffusion gradients and one b0 image. We apply basic dMRI motion-correction  and use probabilistic tractography algorithm from \verb|DiPy| on the deconvolved fODF to estimate the fiber tracks. For tractography, we use the brain mask as the seed region, with density $1$, we use a maximum angle of $75°$ and a stopping criterion threshold of $0.25$ on the FA map and delete short tracks from the output. \\

\noindent\textbf{Evaluation Scores.} We follow Tractometer evaluation scores using Valid Bundles (\textbf{VB}/True Positives), Invalid Bundles (\textbf{IB}/False Positives), Valid Connections (\textbf{VC}/fraction of predicted streamlines part of a valid bundle), Invalid Connections (\textbf{IC}/fraction of predicted streamlines part of an invalid bundle), and No Connections (\textbf{NC}/fraction of predicted streamlines not part of any bundle). \\

\noindent\textbf{Results.} Tractometer results are shown in Table \ref{tab:ismrm_metric}, where our proposed models increase the fraction of valid connections while decreasing the fraction of invalid and non-connected streamlines. This suggests that our models allow for more accurate downstream fiber tracking. While the proposed model does not impact on the number of valid bundles, it halves the number of invalid bundles in a $2$-tissue decomposition setting. We show the partial volume maps on that specific setting on Fig. \ref{fig:pv_ismrm}, where ESD shows a clear separation between the two compartments while CSD overestimates the white matter volume fraction in the isotropic part of the phantom.
A qualitative view of the tractography in the $2$-tissue decomposition setting is presented in figure \ref{fig:tract_ismrm_human}. We see that the streamlines are less noisy with ESD than CSD with better trajectory coherence.

\subsection{Real-world Multi-Shell Human Dataset}
Here, we use the preprocessed multishell human dataset from \cite{tong2020multicenter}. The dataset has three different subjects and we use scans of all three subjects from the first center. The protocol has three $b$-values $\{1000, 2000, 3000\}$ each with $98$ gradients and $27$ $B0$ images. We use a $3$-tissue decomposition for fODF estimation. \\

\noindent\textbf{Evaluation Scores.} Assessing deconvolution performance on a real brain is non-trivial due to the lack of micro-structural ground-truth. Therefore, we follow a downstream utility based evaluation similar to Section \ref{subsec:tractometer}. We estimate the ground-truth partial volume from the T$1$ images of each subject using \verb|FSL FAST| \cite{zhang2001segmentation}. We also analyze qualitative tractograms, computed with  the same probabilistic tractography algorithm as the tractometer dataset. \\

\noindent\textbf{Results.} Quantitative results are shown in Table \ref{tab:kl}. The KL-divergence is consistently improved for the three subjects, suggesting a better estimation of the overall partial volume fractions. Moreover, we show qualitative tractography views of the three subjects in the figure \ref{fig:tract_ismrm_human}. Again, with ESD, we observe less noise and more streamline consistency.

\section{Discussion}
We present an unsupervised rotation-equivariant spherical CNN for the nonlinear estimation of the fiber orientation distribution function from diffusion MRI signals.
With a Cauchy distribution prior on the fODF intensity and increased spherical harmonics order, we obtain sharp and accurate localizations. Our model is flexible and can be used in a wide range of settings, including generic natural image spherical deconvolution. It can work with or without multi-shell data and with an arbitrary number of DWI gradients. Our experiments demonstrate improved micro and macro-structure estimation over state-of-the-art deconvolution frameworks in terms of the detection of smaller crossing-angle fibers, better fiber localization, and better partial-volume estimation. Finally, improved local performances have a positive impact on global white matter fiber tracking in terms of noise and streamline coherence.

\section*{Acknowledgements}
The authors were supported by NIH grants 1R01DA038215-01A1, R01-HD055741-12, 1R01HD088125-01A1,
1R01MH118362-01, R01ES032294, R01MH122447, and 1R34DA050287.

\bibliographystyle{splncs04}
\bibliography{egbib}

\begin{thebibliography}{10}
\providecommand{\url}[1]{\texttt{#1}}
\providecommand{\urlprefix}{URL }
\providecommand{\doi}[1]{https://doi.org/#1}

\bibitem{canales2019sparse}
Canales-Rodr{\'\i}guez, E.J., Legarreta, J.H., Pizzolato, M., Rensonnet, G.,
  Girard, G., et~al.: Sparse wars: A survey and comparative study of spherical
  deconvolution algorithms for diffusion mri. NeuroImage  \textbf{184},
  140--160 (2019)

\bibitem{cohen2016group}
Cohen, T., Welling, M.: Group equivariant convolutional networks. In:
  International conference on machine learning. pp. 2990--2999 (2016)

\bibitem{s.2018spherical}
Cohen, T.S., Geiger, M., Köhler, J., Welling, M.: Spherical {CNN}s. In:
  International Conference on Learning Representations (2018)

\bibitem{daducci2013quantitative}
Daducci, A., Canales-Rodr{\i}guez, E.J., Descoteaux, M., Garyfallidis, E., Gur,
  Y., et~al.: Quantitative comparison of reconstruction methods for intra-voxel
  fiber recovery from diffusion mri. IEEE transactions on medical imaging
  (2013)

\bibitem{gorski1999healpix}
Gorski, K.M., Wandelt, B.D., Hansen, F.K., Hivon, E., Banday, A.J.: The healpix
  primer. arXiv preprint astro-ph/9905275  (1999)

\bibitem{jeurissen2014multi}
Jeurissen, B., Tournier, J.D., Dhollander, T., Connelly, A., Sijbers, J.:
  Multi-tissue constrained spherical deconvolution for improved analysis of
  multi-shell diffusion mri data. NeuroImage  \textbf{103},  411--426 (2014)

\bibitem{karimi2020machine}
Karimi, D., Vasung, L., Jaimes, C., Machado-Rivas, F., Khan, S., et~al.: A
  machine learning-based method for estimating the number and orientations of
  major fascicles in diffusion-weighted magnetic resonance imaging (2020)

\bibitem{lin2019fast}
Lin, Z., Gong, T., Wang, K., Li, Z., He, H., Tong, Q., Yu, F., Zhong, J.: Fast
  learning of fiber orientation distribution function for mr tractography using
  convolutional neural network. Medical physics  \textbf{46}(7),  3101--3116
  (2019)

\bibitem{liu2017multichannel}
Liu, C., Wang, D., Wang, T., Feng, F., Wang, Y.: Multichannel sparse
  deconvolution of seismic data with shearlet--cauchy constrained inversion.
  Journal of Geophysics and Engineering  \textbf{14}(5),  1275--1282 (2017)

\bibitem{lucena2020using}
Lucena, O., Vos, S.B., Vakharia, V., Duncan, J., Ashkan, K., et~al.: Using
  convolution neural networks to learn enhanced fiber orientation distribution
  models from commercially available diffusion magnetic resonance imaging
  (2020)

\bibitem{maier2017challenge}
Maier-Hein, K.H., Neher, P.F., Houde, J.C., C{\^o}t{\'e}, M.A., Garyfallidis,
  E., et~al.: The challenge of mapping the human connectome based on diffusion
  tractography. Nature communications  \textbf{8}(1),  1--13 (2017)

\bibitem{nath2019deep}
Nath, V., Schilling, K.G., Parvathaneni, P., Hansen, C.B., Hainline, A.E.,
  et~al.: Deep learning reveals untapped information for local white-matter
  fiber reconstruction in diffusion-weighted mri. Magnetic resonance imaging
  \textbf{62},  220--227 (2019)

\bibitem{patel2018better}
Patel, K., Groeschel, S., Schultz, T.: Better fiber odfs from suboptimal data
  with autoencoder based regularization. In: International Conference on
  Medical Image Computing and Computer-Assisted Intervention. pp. 55--62.
  Springer (2018)

\bibitem{perraudin2019deepsphere}
Perraudin, N., Defferrard, M., Kacprzak, T., Sgier, R.: Deepsphere: Efficient
  spherical convolutional neural network with healpix sampling for cosmological
  applications. Astronomy and Computing  \textbf{27},  130--146 (2019)

\bibitem{sedlar2020diffusion}
Sedlar, S., Papadopoulo, T., Deriche, R., Deslauriers-Gauthier, S.: Diffusion
  mri fiber orientation distribution function estimation using voxel-wise
  spherical u-net. In: Computational Diffusion MRI, MICCAI Workshop (2020)

\bibitem{tong2020multicenter}
Tong, Q., He, H., Gong, T., Li, C., et~al.: Multicenter dataset of multi-shell
  diffusion mri in healthy traveling adults with identical settings. Scientific
  Data  (2020)

\bibitem{tournier2007robust}
Tournier, J.D., Calamante, F., Connelly, A.: Robust determination of the fibre
  orientation distribution in diffusion mri: non-negativity constrained
  super-resolved spherical deconvolution. Neuroimage  \textbf{35}(4),
  1459--1472 (2007)

\bibitem{tournier2019mrtrix3}
Tournier, J.D., Smith, R., Raffelt, D., Tabbara, R., Dhollander, T., et~al.:
  Mrtrix3: A fast, flexible and open software framework for medical image
  processing and visualisation. NeuroImage  \textbf{202},  116137 (2019)

\bibitem{wedeen2000mapping}
Wedeen, V., Reese, T., Tuch, D., Weigel, M., Dou, J., Weiskoff, R., Chessler,
  D.: Mapping fiber orientation spectra in cerebral white matter with
  fourier-transform diffusion mri. In: Proceedings of the 8th Annual Meeting of
  ISMRM (2000)

\bibitem{yan2018estimating}
Yan, H., Carmichael, O., Paul, D., Peng, J., Initiative, A.D.N., et~al.:
  Estimating fiber orientation distribution from diffusion mri with spherical
  needlets. Medical image analysis  \textbf{46},  57--72 (2018)

\bibitem{zhang2001segmentation}
Zhang, Y., Brady, M., Smith, S.: Segmentation of brain mr images through a
  hidden markov random field model and the expectation-maximization algorithm.
  IEEE transactions on medical imaging  \textbf{20}(1),  45--57 (2001)

\end{thebibliography}

\end{document}